

\magnification=\magstep1
\baselineskip=20 pt plus 1pt minus 1pt
\vsize=8.9truein
\hsize=6.8truein
\hfuzz=1.5pt
\footline={\ifnum\pageno=1 \hfil\else\centerline{\folio}\fi}
\def\v{\vskip6pt}
\def\no{\noindent}
\def\h{\textstyle {1 \over 2}}
\def\ad#1{{(\hbox{ad$\,$\bf #1})}}
\def\ads#1{{(\hbox{ad$\,${\smbold #1}})}}
\def\bn#1{{\bf #1}}
\def\ft#1{$\,$[#1]}
\def\p{\partial}
\def\E#1{Eq.~(#1)}
\def\Es#1{Eqs.~(#1)}

\newbox\dstrutbox
\setbox\dstrutbox=\hbox{\vrule height10.5pt depth5.5pt width0pt}
\def\dstrut{\relax\ifmmode\copy\dstrutbox\else\unhcopy\dstrutbox\fi}
\def\tablerule{\noalign{\hrule}\cr}

\font\smchapt=cmbx12

\font\smbold=cmbx8
\font\reglar=cmr10
\centerline{\smchapt The Robinson-Trautman Type III Prolongation
Structure Contains K${}_{\bf 2}$}
\reglar\v\v\baselineskip=14pt
\centerline{J. D. Finley, III}
\v
\centerline{Department of Physics and Astronomy}\par
\centerline{University of New Mexico}\par
\centerline{Albuquerque, N.M.  87131}
\rightline{Internet:~~finley@tagore.phys.unm.edu}\par
\rightline{FAX:~~(505)-277-1520\phantom{Internet,Internet}}
\vskip2truein
{\narrower  The minimal prolongation structure for the Robinson-Trautman
equations
of Petrov type III is shown to always include the infinite-dimensional,
contragredient algebra, $K_2$, which is of infinite growth.
Knowledge of faithful representations of this algebra would allow the
determination of B\"acklund transformations to evolve new solutions.
\medskip}
\vskip1.4truein
\rightline{Running Title:~~~ Robinson-Trautman Type III and K$_2$}
\rightline{MCS Numbers:~~ 17B65, 83C20}
\vfill\eject
\reglar
\baselineskip=16pt plus 2pt minus 2pt
\reglar
The general class of Robinson-Trautman solutions\ft{1} to the vacuum
Einstein field equations have been important examples of exact solutions
for many years, albeit they seem to have various difficulties with respect
to their interpretation\ft{2}.  They are solutions characterized by having a
repeated principal null direction, which is of course geodesic and shearfree,
and is required to be diverging but not twisting.  The standard
reference\ft{3} gives the general form of the metric which any Einstein space
must have if it
permits such a repeated principal null direction, and notes that all possible
algebraically-special Petrov types are allowed.  In the case of Petrov
type III, the field equations are\ft{3} first reduced to
$$K = \Delta \log P \equiv 2
P^2\p_{\lower1.5pt\hbox{$\scriptstyle\zeta$}}\p_{\,\overline{\zeta}}\log P =
-3[f(\zeta,u) +\overline{f}(\overline{\zeta},u)] \quad,\eqno(1)$$
\no where $K$ is the Gaussian curvature of the 2-surface spanned by
$\zeta$ and $\overline{\zeta}$.  This equation
determines the general RT-solution of Petrov type III.  However,
since $u$ is nowhere explicitly mentioned within the partial differential
equation
(pde), it is well-known\ft{3} that one could always simply ignore that
dependence, perform a coordinate transformation sending $f(\zeta)
\rightarrow\zeta$, leaving the curvature completely invariant, and
reducing our equation to the rather simple-appearing equation\par
{\abovedisplayskip=6pt plus 3pt minus 5pt\vskip-8truept
$$\eqalignno{K \quad = \quad 2P^2
\p_{\lower1.5pt\hbox{$\scriptstyle\zeta$}}\p_{\,\overline{\zeta}}
\log P & = -{\textstyle{3\over 2}}(\zeta + \overline{\zeta})\quad,\cr
\hbox{or}\qquad u_{xy} & = \h(x+y)e^{-2u}\quad,\qquad
\hbox{where}\ \ \log P \equiv u\;,\qquad&(2)\cr}$$}
\no the subscripts denote partial derivatives, and the symbols $\{x,y\}$
have been introduced instead of $\{
\zeta,\overline{\zeta}\}$, both to simplify the typography and to
normalize the equation so that the coefficient has a value
which will prove convenient.
\v
As already pointed out, all Petrov type III solutions of the vacuum field
equations with diverging, non-twisting null directions are
determined by the general
solution of \E{2}.  Nonetheless, only one rather trivial solution is
available for study, namely $P = (\zeta+\overline{\zeta})^{3/2}$,
even though all its Lie symmetries have been found.\ft{4}  This
unfortunate situation has caused us to apply the general methods of
Estabrook and Wahlquist\ft{5} to this equation, for determination of
(pseudo)-potentials, in the hope of generating new solutions.
The EW procedure is a particular approach to the determination of
non-local symmetries of a pde\ft{6}. It has
been used successfully\ft{7-10} in many contexts, although it must be
admitted that applications toward finding RT solutions of Petrov type
II were unsuccessful\ft{11-12}.
\v
The situation for Petrov type III seems to be much better.
We have in fact obtained a complete, but so far still abstract,
description of the space of allowed pseudopotentials.
The unexpected consequence of this search was that the smallest space
that allows a solution of the problem must be
a carrier space of a realization, via vector fields,
of an algebra of infinite growth\ft{13,14}, usually referred to as
$K_2$\ft{14,15}.  This algebra has so far resisted any attempts to find
explicit realizations.  The name was created by Kac\ft{14} in his early
article separating those algebras now called Kac-Moody algebras away from
classes of much larger algebras.  Kac used $K_2$ as a ``simple'' example
of a contragredient algebra not in the Kac-Moody class.
\v
We begin our investigation by describing the solution space for the
equation as a surface,
$Y$, in the jet space, $J^{(2)}$, that treats dependent functions and their
first and second derivatives as independent quantities until a specific
solution is obtained.  The Estabrook-Wahlquist procedure guides one in
searches for $\bf F$ and $\bf G$, vector fields over a space, $W$, of
pseudopotentials,
$\{w^A\mid A=1,\,\ldots\,,\,N\}$, that we wish to adjoin to the original
jet space.  These vector fields provide prolongations of the (usual) total
derivative operators on the jet space, {\it i.e.,} $\{D_x,\,D_y\}$, to
the combined space of variables, $J\oplus W$, which must then
must satisfy the zero-curvature equations when restricted to $Y$:
$$\left[\tilde D_x + {\bf F}\,,\,\tilde D_y + {\bf G}\right] = 0\quad,
\eqno(3)$$
\no where the tilde indicates that the operators have been restricted to
the subspace, $Y$, of solutions.
This is a slight generalization of the usual notion of the zero-curvature
equations of Lax or of Zakharov and Shabat\ft{16}, since
$\bf F$ and $\bf G$ are simply elements of an {\bf abstract} Lie algebra,
of vector fields, with neither the coordinates, nor even $N$, yet determined.
\v
Following the approach of Cartan\ft{17}, the EW procedure for a pde in
2 independent variables may be described as follows\ft{18}.
We first choose a (closed) ideal, ${\cal K}$, generated by a set of 2-forms,
$\{\alpha^r\}$, that describes the original pde.
We then adjoin the variables $w^A$ to the system by
appending (to the original ideal) contact forms, $\omega^A$,
for each of these new variables,
and insisting that the ideal remain closed:
$$\eqalign{\omega^A  =  & - dw^A + F^A dx + G^A dy, \quad\cr
dF^A \wedge dx + & dG^A\wedge dy = f^A\!\!{}_r\, \alpha^r + \eta^A{}_B
\wedge \omega^B\quad,\cr}\qquad  A = 1, \ldots,
N\quad,\eqno(4)$$
where the functions $F^A$ and $G^A$ are the coefficients of the
vector fields, $\bf F$ and $\bf G$, that define the zero-curvature
representation of the problem. These brief sentences
describe the ``essence'' of the EW procedure, which embodies two notions,
the first being the choice of a sufficiently-small ideal that calculations
can be carried out successfully, while the other is that the new potentials
are all allowed, from the beginning, to depend on each other, thereby
rendering the process of discovering them nonlinear, and justifying the
name, ``pseudopotentials,'' for these additional variables.
The newly-introduced functions $f^A{}_r$ and 1-forms $\eta^A{}_B$
constitute ``Lagrange
multipliers'' for the system, their existence being the explicit
characterization of closure of the prolonged ideal.  Because of this
the final choices of
$\bf F$ and $\bf G$ must maintain {\bf non-zero} the multipliers
$f^A{}_r$, since they retain the information needed by the procedure to
``remember'' the original ideal, and therefore the given pde.
\v
Comparison of the coefficients of the various independent 2-forms
on both sides of \E{4} determines those jet-variables on which $F^A$ and
$G^B$ do {\bf not} depend, {\bf and} expresses the Lagrange multipliers,
in terms of derivatives of the $F^A$ and $G^B$.  The only remaining
requirements of closure, in
\E{4}, are that the coefficient of $dx\wedge dy$ should vanish.  This
particular coefficient is the expression of the commutator in \E{3}, obtained
by this method.
\v For the RT equation, we choose our ideal, $\cal K$, as that ideal
within $\Lambda^2(J^{(1)})$ generated by
$$(du-p\,dx)\wedge dy\quad,\quad (du-q\,dy)\wedge dx\quad,\quad dp\wedge dx
- dq\wedge dy+(x+y)e^{-2u}\,dx\wedge dy\quad.\eqno(5)$$
\no While this is in fact not the smallest choice, its symmetry makes the
problem rather easier.  (We will show in Appendix I that making
other choices does not change our (minimal) result, concerning $K_2$.)
Comparing coefficients gives us the non-dependencies, the Lagrange multipliers,
and the commutator equation for this particular ideal:
$$\eqalign{{\bf F}_q = 0 = {\bf G}_p\quad; & \qquad
\lambda_1^A = G_u^A\quad,\quad\lambda_2^A=F_u^A\quad,\quad
\lambda_3^A=Z^A\quad,\cr
[{\bf F} + \p_x, {\bf G} + \p_y] &  = -p\,{\bf G}_u+q\,{\bf F}_u +
\h(x+y)e^{-2u}
({\bf F}_p-{\bf G}_q)\quad.\cr}\eqno(6)$$
\v Comparing coefficients in \E{6}, the stated dependencies
allow us to infer the existence of
vertical vector fields $\bf B$, $\bf C$, and $\bf Z$ such that
$${\bf F} = p\,{\bf Z} + {\bf B}\quad,\quad{\bf G} = - q \,{\bf Z} + {\bf
C}\quad,\quad
{\bf Z}_u = 0\quad,\eqno(7)$$
\no Re-inserting these forms into \E{6}, it becomes a polynomial
in $p$ and $q$,
so that the vanishing of all of the separate coefficients gives the
following equations:
$$\eqalignno{[{\bf Z}\,,\,{\bf C}] = -{\bf C}_u + {\bf Z}_y\quad,&\quad
[{\bf Z}\,,\,{\bf B}] = +{\bf B}_u + {\bf Z}_x\quad,&(8a)\cr
[{\bf B}\,,\,{\bf C}] = {\bf B}_y-{\bf C}_x & +
(x+y)e^{-2u}{\bf Z}\quad.&(8b)\cr}$$
\v To proceed further with the integration of these equations,
we must make {\it some} assumption concerning the dependence on the
independent variables.  In most studies of pde's via the EW prolongation
procedure, it is common to assume no dependence on the
independent variables\ft{19}, although the Ernst equation has indeed
been an exception\ft{10} to this.  However, having explicit dependence
on those variables, it should be clear that this equation will require some
dependence of $\bf F$ and $\bf G$ on $\{x,y\}$.  Originally, we
argued that the most reasonable approach would be to assume that ${\bf F}_x =
0 = {\bf G}_y$, since those derivatives would not appear in the
final expression anyway.  This approach can in fact be completed, and
will be discussed in Appendix II;  however, at least in this instance, we
will show that it is gauge equivalent to the opposite approach, namely that
${\bf F}_y = 0 = {\bf G}_x$, which is the rather simpler
road we shall now follow.  Since
$\bf Z$ appears in the expressions for both of $\bf F$ and $\bf G$,
this requires that ${\bf Z}_x = 0 = {\bf Z}_y$, and reduces \Es{8} to
$$[{\bf Z}\,,\,{\bf C}] = -{\bf C}_u \quad,\quad
[{\bf Z}\,,\,{\bf B}] = +{\bf B}_u \quad,\quad
[{\bf B}\,,\,{\bf C}] =  (x+y)e^{-2u}\>{\bf Z}\quad,\eqno(9)$$
\no where  ${\bf B = B}(x)$ and ${\bf C = C}(y)$.
\v
The first two of \Es{9} are simply flow equations for a vector
field\ft{18}, which are immediately integrated to give
$${\bf B}(x,u) = e^{+u\ads{Z}}\,{\bf R}(x)\qquad,\qquad {\bf C}(y,u) =
e^{-u\ads{Z}}\,{\bf S}(y)\quad,\eqno(10)$$
\no where we have indicated explicitly the assumed
$x$- and $y$-dependence.  Inserted into the last of
\Es{9}, these forms give us the ``last'' requirement,
$$[ e^{+u\ads{Z}}\,{\bf R}(x)\,,e^{-u\ads{Z}}\,{\bf S}(y)]
\quad=\quad (x+y)e^{-2u}\>{\bf Z}\quad.\eqno(11)$$
\no We interpret this condition as the agreement of two power series
in $u$; the coefficients of $u^k/k!$ are given by
$$\sum_{m=0}^k {-1\overwithdelims() 2}^k{k\choose m}[{\bf R}_{(k-m)}\,,
{\bf S}_{(m)}] = (x+y)\,{\bf Z}\quad,\;\forall k = 0,1,2,\,\ldots\quad,
\eqno(12)$$
\no where the subscripts in parentheses indicate repeated
commutators with $\bf Z$:
$$ {\bf R}_{(m)} \equiv (-1)^m\ad{Z}^m\,{\bf R}(x)\quad,\quad {\bf S}_{(m)}
\equiv \ad{Z}^m\,{\bf S}(y)\quad,\;\forall m = 0,1,2,\,\ldots\quad.
\eqno(13)$$
\v
The 0-th order term of \E{11} implies that
$[{\bf Z}\,,[{\bf R}(x)\,,{\bf S}(y)]] = 0$.   Inserting this fact
into the Jacobi identity shows that the commutators
$[{\bf R}_{(k-m)}\,,{\bf S}_{(m)}]$ are actually
independent of the value of $m$,
allowing us to sum the series in \Es{12}:
$$[{\bf R}_{(i)}\,,{\bf S}_{(j)}] = (x+y)\,{\bf Z}\quad,\;\forall i,j =
0,1,2,\,\ldots\quad.\eqno(14)$$
\no While there is no obvious requirement that the
various ${\bf R}_{(i)}$, for example, be parallel, the fact that the
right-hand side of the equation depends on neither $i$ nor $j$ does
surely suggest such a thought.  In Appendix II, we describe the more
general case, while here we take as an additional assumption that
all ${\bf R}_{(i)}$ are parallel, and that all ${\bf S}_{(j)}$ are parallel.
The coefficients of proportionality are determined uniquely,
causing the infinite sums for ${\bf B}$ and $\bf C$
to both become proportional to
$e^{-u}$.  This allows all $u$-dependence to be factored out of \E{11},
reducing it to the simpler requirement:
$${\bf [ Z,\, S] = S\quad,\quad [Z,\,R] = - R\quad,
\quad \Longrightarrow\quad
[R(x),\,S(y)]} = (x+y)\,{\bf Z}\quad.\eqno(15)$$
\no \v
The explicit existence of $x$ and $y$ in the original pde generated
the need for $x$- and $y$-dependence of our prolongation vector fields.
Since, however, they are linear in those variables, and display themselves
explicitly that way in the last of the equations in \Es{15},
it seems sufficient to consider the yet-further special case when
${\bf R}(x)$ and ${\bf S}(y)$ are just first-order polynomials in their
respective jet variables:
$${\bf R}(x) \equiv -{\bf f}_1 - x{\bf f}_2\qquad,\qquad{\bf S}(y)\equiv
+{\bf e}_2 + y{\bf e}_1\quad.\eqno(16)$$
\no Inserting these forms into \Es{15} gives the
complete presentation of the prolongations of the total derivatives as
explicit functions of the original jet variables:
$${\bf F} = p{\bf Z} - e^{-u}({\bf f}_1 + x{\bf f}_2)\quad,\quad
{\bf G} = -p{\bf Z} + e^{-u}({\bf e}_2 + y{\bf e}_1)\quad.\eqno(17)$$
\no The final
requirements on the system are simply statements of some of the
commutators of the vector fields on the fibers (of pseudopotentials)
themselves:
$$\eqalign{[{\bf Z},{\bf e}_i]={\bf e}_i \quad,\quad[{\bf Z},{\bf f}_i]
=-{\bf f}_i\quad,\;i=1,2\quad,\cr
[{\bf e}_2,{\bf f}_1] = 0 = [{\bf e}_1,{\bf f}_2]\quad,
\quad [{\bf e}_1,{\bf f}_1] = {\bf Z} =
[{\bf e}_2,{\bf f}_2]\quad.\cr}\eqno(18)$$
\v Referring to \Es{6}, we see that the three Lagrange multipliers are now
proportional to $\{{\bf R}(x), {\bf S}(y), {\bf Z}\}$.  Our next task is to
determine a realization of the algebra defined by these 5 generators,
which maintains these three quantities linearly independent.
This algebra, defined by the 5 generators above, is still not completely
displayed since the quantities $[{\bf e}_1, {\bf e}_2]$ and $[{\bf f}_1, {\bf
f}_2]$
are not given, and are therefore to be considered arbitrary modulo the
requirements of the Jacobi identity.  As examples of these sorts of
requirements, it is straight-forward to show
that
$$\eqalign{[{\bf Z},\{\ad{${\bf e}_1$}^n{\bf e}_2\}] & = (n+1)
\{\ad{${\bf e}_1$}^n{\bf e}_2\}\;,\cr
[{\bf e}_2,\{\ad{${\bf f}_2$}{}^{m+1}{\bf f}_1\}]
& = -(m+1)\{\ad{${\bf f}_2$}^m{\bf f}_1\}\;.\cr}\eqno(19)$$
\no A presentation of any Lie algebra as a direct sum of
subspaces\ft{14,20},
with the following requirement on the Lie bracket operation is referred
to as an (integer)-graded Lie algebra:
$${\cal G} = \mathop{\oplus}_{i=-\infty}^\infty{\cal G}_i\;,\qquad
[{\cal G}_i\,,\,{\cal G}_j] \subseteq {\cal G}_{i+j}\quad.\eqno(20)$$
\no If $d_i$ is the dimension of ${\cal G}_i$, as a vector space, then
$$r\equiv \overline{\lim_{i\rightarrow\infty}}\left\{{\log\left(
\sum\limits_{j=-i}^i d_j\right)\Biggm/\log(i)}\right\}\eqno(21)$$
\no is called\ft{13,20} the
{\it growth} of the full Lie algebra, ${\cal G}$.  Finite-dimensional
Lie algebras have growth 0, while those usually referred to as
Kac-Moody algebras have finite growth.  We refer to
$\hat{{\cal G}} \equiv {\cal G}_{-1}\oplus{\cal G}_0\oplus{\cal G}_1$
as {\it the local part} of ${\cal G}$, and supplement our definition of
a graded algebra by insisting that it should be (algebraically)
generated by commutators of its local part, so that the grading is
then well-defined.  (This is of course what one would expect.)
For our algebra, we take the local part as follows:
$${\cal G}_0 \equiv \{{\bf Z}\}\quad,\quad {\cal G}_{-1}\equiv \{{\bf f}_i\mid
i=1,2\}\quad,\quad
{\cal G}_1\equiv \{{\bf e}_i\mid i=1,2\}\quad.\eqno(22)$$
The first equality in
\Es{19} then tells us that the dimension of ${\cal G}_i$ is growing rapidly
{\bf unless} the objects $\{\ad{${\bf e}_1$}^n{\bf e}_2\}$
were to vanish from some value of $n$ onward, which is indeed what occurs
in a Kac-Moody algebra.  The second equality in \Es{19}
tells us that if those quantities were to vanish, there would be a downward
cascade causing that particular entire part of the structure to vanish,
leaving us with zero values for our Lagrange multipliers, which is of course
unacceptable.  We may therefore conclude that this algebra does
indeed grow quite fast.
\v
In fact this algebra may be completely identified.  It is the simplest
contragredient algebra of infinite growth, referred to as $K_2$.
In Chapter II of Ref. 14, Kac defines general contragredient algebras
associated with a given matrix $A$ with integer elements.
They are integer-graded algebras with certain requirements on the
Lie brackets of the basis elements of the local part.  Let $\{f_i\}$,
$\{h_i\}$, and $\{e_i\}$, be basis vectors for ${\cal G}_{-1}$,
${\cal G}_0$, and ${\cal G}_1$, respectively.  We first require that
their commutators satisfy the following:
$$[{\bf e}_i,{\bf f}_j] = \delta_{ij}{\bf h}_i\quad,\quad [{\bf h}_i,
{\bf h}_j] = 0\quad,\quad [{\bf h}_i,{\bf e}_j] =
A_{ij}{\bf e}_j\quad,\quad [{\bf h}_i,{\bf f}_j] =
-A_{ij}{\bf f}_j\eqno(23)$$
The contragredient Lie algebra is then the minimal graded Lie algebra,
with local part $\hat{{\cal G}}$.  (Beginning with any algebra generated by
this local part, finding the largest
homogeneous ideal that contains no elements of ${\cal G}_0$ (except 0) and
then factorizing the algebra over this ideal will create the minimal one.)
In the special case that the matrix $A$ has its diagonal
elements positive (usually normalized to +2), its
off-diagonal elements non-positive, and all $A_{ij} = 0 \Leftrightarrow
A_{ji} = 0$, for $i\ne j$, then it is called a generalized Cartan matrix.
\v
For our problem, we may now consider a contragredient algebra
with matrix $A$ such that
$$A = \pmatrix{1&1\cr1&1\cr}\quad.\eqno(24)$$
\no Our 3 sets of basis vectors are $\{h_1, h_2\}$,
$\{e_1, e_2\}$, and $\{f_1, f_2\}$, so that
$[h_i,e_j] = e_j$, etc., for $i,j = 1,2$.  Therefore, we see that
$(h_1-h_2)$ is central, so that we may factor our algebra by it.  The
resulting algebra is $K_2$, except that Kac normalizes his vectors so
that all the elements of $A$ have the value 2 instead of 1.
$K_2$ is now easily seen to be isomorphic to the
prolongation algebra we have determined for the RT equation of type III,
with $Z\rightarrow h_1\pmod{h_1-h_2}$.
\v
Having determined the smallest prolongation algebra, the next step in
the process of finding new solutions is to write down explicit vector-
field (or matrix) realizations of this algebra, use the variables in
the carrier space as pseudopotentials, pick out a B\"acklund
transformation, take the one existing solution, and begin to generate
new ones, as has been done many times before with many other interesting
pde's.  The difficulty, in this case, is that {\bf no} realizations of
$K_2$ have yet been discovered.  Since this is indeed the minimal
prolongation algebra, we see that there is considerable correspondence
between the two problems.  It seems reasonable to suppose that finding
new solutions is equivalent to evolving realizations of this algebra.
Therefore, the main purpose of this report is to encourage its
listeners to try to achieve at least some non-trivial realization of $K_2$.
\v
\par \no{{\it Acknowledgments:}}~~Considerable appreciation needs to be
expressed for discussions with Mikhail Saveliev, who noticed the gauge
transformation described in Appendix II, to Mark
Hickman, who suggested that this problem should be amenable to prolongation
techniques, and to John K. McIver, who listened to many descriptions of the
needs for this problem.
\vfill\eject
\centerline{{\bf Appendix I:~~Other Choices of Generating Ideal}}\v
We have found it convenient to use a symmetric choice for the
generators of our subideal of the complete, restricted contact module.
Other particular choices of ideal may generate
distinct maximal algebras\ft{21,22}; nonetheless we now show that other
plausible choices do not actually change the minimal algebra involved in
the prolongation process for this pde. The sine-Gordon equation
is very similar to our equation, simply not involving explicitly
the independent variables.  Our symmetric ideal is
actually modelled on that used by Shadwick\ft{23} for the sine-Gordon
equation.  On the other hand, many other authors, including in particular
Hoenselaers\ft{8}, have used an ideal for the sine-Gordon equation that
is asymmetric, {\bf and} contains fewer generators. These two
actually constitute all the reasonable choices one can make\ft{24}.
\v
Following Hoenselaers' model\ft{8}, an alternative ideal
would have the following generators:
 $$(du-p\,dx)\wedge dy\quad,\quad dp\wedge dx +\h(x+y)e^{-2u}\,dx\wedge
dy\quad.\eqno(A1.1)$$
\no Since this does have both fewer generators and fewer variables, not
using $q$, than the one we described in
\Es{6}, one could indeed hope for ``nicer'' results.  Following the same
procedure as before, the analogue of \Es{7} is quickly found to be
$$\eqalign{{\bf F}={\bf F}(x,y,p)\;,\ {\bf G} & ={\bf G}(x,y,u)\quad; \qquad
\lambda_1 = {\bf G}_u\quad,\quad\lambda_2={\bf F}_p\quad,\cr
[{\bf F} + \p_x, {\bf G} + \p_y] &  = -p\,{\bf G}_u + \h(x+y)e^{-2u}
\,{\bf F}_p\quad.\cr}\eqno(A1.2)$$
\no Introducing the new quantity ${\bf P} \equiv {\bf G} + \h{\bf G}_u $,
reduces the commutator equation in \Es{A1.2} to the simpler form $[{\bf F} +
\p_x\,,{\bf P} + \p_y] = -p\,{\bf P}_u$.  The general solution
can be worked out in an analogous fashion to that shown in
Ref. 22; however,  by first taking two successive derivatives
with respect to $p$, resulting
in $[{\bf F}_{pp}\,,{\bf P}] = 0$, we can pick out the smallest interesting
piece of it, again in a manner analogous to Ref. 8, by simply
setting both the
objects in this last commutator,
separately, to zero, which gives us
$$ {\bf F} = p{\bf A} + {\bf B}\quad,\qquad {\bf G} = e^{-2u}{\bf C}\quad,
\eqno(A1.3)$$
\no where $\{\bf A, B, C\}$ all depend on both $x$ and $y$, but with
no necessity for dependence more complicated than linear.  Inserting
these forms back into the original commutator
equation produces the following
results, with the $u$-dependence already completely satisfied:
$${\bf A}_y = 0 = {\bf B}_y\ ;\quad [{\bf A}\,,{\bf C}] = 2{\bf C}\ ;\quad
{\bf C}_x + [{\bf B}\,,{\bf C}] = \h\,(x+y)\,{\bf A}\quad.\eqno(A1.4)$$
\no Expanding each of our these vector fields as first-order polynomials,
 in the form
$${\bf A} = 2{\bf A}_0 + 2x{\bf A}_1\;,\ {\bf B} = {\bf B}_0 + x{\bf B}_1\;,\
{\bf C} = {\bf C}_0 + x{\bf C}_1 + y{\bf C}_2\quad,\eqno(A1.5)$$
\no we easily calculate the commutators required, and find that
linear independence of our Lagrange multipliers insists that
the set $\{{\bf A}_0,{\bf B}_0,{\bf C}_0, {\bf C}_2\}$ must remain linearly
independent, and of course non-zero.  There are two plausible
special cases of interest here:
Case 1 sets ${\bf A}_1 = 0 = {\bf B}_1$, while case 2 sets ${\bf A}_1 = 0 =
{\bf C}_1$.  For case 1, we have 5 generators, with all but 4 of the 10
commutator products already determined.  We present the commutators in
the form of a table, and do not bother to indicate the lower-triangular
portion since it is of course skew-symmetric:
$${\vbox{\offinterlineskip \halign{ &\vrule#& \dstrut\hfill\ #$\>$\hfill\cr
\omit&\omit&\multispan{11}\hrulefill\cr
 \omit &\omit &width2pt& ${\bf A}_0$ && ${\bf B}_0$ &&${\bf C}_0$
 &&${\bf C}_1$ &&${\bf C}_2$&\cr
\noalign{\hrule height2pt}\cr
& ${\bf A}_0$ &width2pt&0&&\omit&&${\bf C}_0$&& ${\bf C}_1$&& ${\bf C}_2$ &\cr
\tablerule
& ${\bf B}_0$ &width2pt& \omit && $0$ &&$-{\bf C}_1$&&${\bf A}_0$&&${\bf
A}_0$&\cr
\tablerule
&${\bf C}_0$ &width2pt&\omit&&\omit&&0&&\omit&&\omit&\cr
\tablerule
&${\bf C}_1$ &width2pt&\omit&&\omit&&\omit&&0&&\omit&\cr
\tablerule
&${\bf C}_2$ &width2pt&\omit&&\omit&&\omit&&\omit&&0&\cr
\noalign{\hrule} \cr }}}\eqno(A1.6)$$
\no The 4 omitted entries in the upper-triangular portion must still be
determined.  Taking ${\bf A}_0$ as an element in the Cartan
subalgebra, ${\cal G}_0$, we see that all of ${\bf C}_i$ constitute positive
roots, but there are no immediately-determined negative roots.  A plausible
``cure'' for this is to identify the undetermined commutator, $[{\bf A}_0\,,
{\bf B}_0]$ as a negative root; i.e., to assume that
$[{\bf A}_0\,,[{\bf A}_0\,,
{\bf B}_0]] = -[{\bf A}_0\,,{\bf B}_0]$, consistent with the
Jacobi identity.  Another identification that reduces the number of
unknown commutators in a manner consistent with the Jacobi identity is
to identify ${\bf C}_1 = {\bf C}_2$; this has the obvious justification that
it causes ${\bf C}$ to depend only on $x+y$, just as does the pde itself.
At this point, the entire algebra---with the notable exception of
${\bf B}_0$---can be identified with (our version of) $K_2$ again, using
the contragredience matrix $A$ given by \E{24}:
$${\bf A}_0 \rightarrow {\bf h}\,,\ {\bf C}_0 \rightarrow {\bf e}_1
\,,\ {\bf C}_1 = {\bf C}_2 \rightarrow {\bf e}_2\,,\ [{\bf A}_0\,,{\bf B}_0]
\rightarrow {\bf f}_2\,,\ [{\bf B}_0\,,[{\bf A}_0\,,{\bf B}_0]]\rightarrow
{\bf f}_1\,.\eqno(A1.7)$$
\no Since ${\bf B}_0$ is a subalgebra, we may then identify the algebra at
this point with the semi-direct sum of $K_2$ and $\{{\bf B}_0\}$, which is
satisfactory for our current purposes.  (See the related result in
Appendix II.)
\v
Following case 2 equally far, $\bf F$ depends only on $x$
(and $p$) while $\bf G$ depends only on $y$ (and $u$), as was the
case for the symmetric ideal already discussed.  Again we have five
generators with only six of the commutators already determined.
The known commutators are
$${\vbox{\offinterlineskip \halign{ &\vrule#& \dstrut\hfill\ #$\>$\hfill\cr
\omit&\omit&\multispan{11}\hrulefill\cr
 \omit &\omit &width2pt& ${\bf A}_0$ && ${\bf B}_0$ &&${\bf B}_1$
 &&${\bf C}_0$ &&${\bf C}_2$&\cr
\noalign{\hrule height2pt}\cr
& ${\bf A}_0$ &width2pt&0&&\omit&&\omit&& ${\bf C}_0$&& ${\bf C}_2$ &\cr
\tablerule
& ${\bf B}_0$ &width2pt& \omit && $0$ &&\omit&&0&&${\bf A}_0$&\cr
\tablerule
&${\bf B}_1$ &width2pt&\omit&&\omit&&0&&${\bf A}_0$&&0&\cr
\tablerule
&${\bf C}_0$ &width2pt&\omit&&\omit&&\omit&&0&&\omit&\cr
\tablerule
&${\bf C}_2$ &width2pt&\omit&&\omit&&\omit&&\omit&&0&\cr
\noalign{\hrule} \cr }}}\eqno(A1.8)$$
\no The 4 omitted entries in the upper-triangular portion must still be
determined.  Taking ${\bf A}_0$ as an element in the Cartan subalgebra,
${\cal G}_0$, we see that the ${\bf C}_i$ constitute positive
roots, but there are no immediately-determined negative roots.  A plausible
``cure'' identifies the ${\bf B}_j$ as negative roots, thereby
determining two of the previously-unknown commutators.  This is consistent
with the Jacobi identity and directly identifies the algebra as $K_2$, with
matrix $A$ given by \E{24}:
$${\bf A}_0 \rightarrow {\bf h}\,,\ {\bf B}_0\rightarrow
-{\bf f}_2\,,\ {\bf B}_1\rightarrow -{\bf f}_1\,,
{\bf C}_0 \rightarrow {\bf e}_1\,,\ {\bf C}_2 \rightarrow
{\bf e}_2\,.\eqno(A1.9)$$
\no Therefore, this ideal also always leads to algebras
of infinite growth, certainly containing $K_2$.
\vfill\eject
\centerline{{\bf Appendix II:~~Other Choices of $\{x,y\}$-dependence}}\v
 Following \Es{9}, we considered further only the special case
where ${\bf F}_y = 0 = {\bf G}_x$, resulting in \Es{10}, which were much
simpler.
 However, the alternative set of
assumptions is also viable and justifiable, i.e., setting ${\bf F}_x = 0 =
{\bf G}_y$.  We argue that this is reasonable since these particular
derivatives never appear, explicitly, within \Es{9}.  At this point
the first two of \Es{9} each have the form of a vector-field-valued
pde which is somewhat more complicated than simply the usual flow equations.
However, since the derivative operator acting on either one of the
(to-be-determined) vector fields gives exactly zero when operating on
the other---due to the assumptions just made---we can actually still
manage to integrate these equations, the solution to which we describe
below in the following terms.
\v\goodbreak
\no{\bf Lemma:}~~Solution of the pde $\bigl[{\bf A},\,{\bf R}\bigr] =
{\bf A}_x + {\bf R}_u$.\par
We suppose given two vertical vector fields, ${\bf A}$ and ${\bf R}$,
elements of the Lie algebra of vector fields over the space $W$ of our
pseudopotentials.  As these lie in the tangent bundle to fibers over
$J^{(2)}$, they also depend on two {\bf disjoint\/} sets of
other variables, say ${\bf A} = {\bf A}(x,y)$
and ${\bf R} = {\bf R}(u,v)$, and are required to satisfy the pde
$$\bigl[{\bf A},\,{\bf R}\bigr] = {\bf A}_x + {\bf R}_u\quad,\eqno(A2.1)$$
\no where as usual the subscripts indicate partial derivatives.
The solution is determined by first differentiating the equation
with respect to, say, $x$, which annuls the derivative of ${\bf R}$, providing
a flow equation for ${\bf A}_{xx}$ along ${\bf R}$, which we integrate,
taking proper care of the fact that while the general form of ${\bf R}$
depends on both $u$ and $v$, ${\bf A}_{xx}$ depends on neither one.
We then differentiate with respect to $u$, and follow an analogous procedure
for ${\bf R}_{uu}$. The general solution is then obtained by substituting
back in to the original equation and making all ``constants of integration''
behave properly.
That solution is determined by the sets of vector fields
${\bf A}_0(y) \equiv \sum_{m=0}^\infty {y^m\over m!}{\bf A}_{0m}$ and
${\bf R}_0(v) = \sum_{k=0}^\infty {v^k\over k!}\, {\bf R}_{0m}$,
and the field ${\bf A}_{10}$ or ${\bf R}_{10}$, which are related
symmetrically by ${\bf A}_{10} - {\bf R}_{10} = [{\bf R}_{00}\,,{\bf
A}_{00}]$, such that
$$\eqalign{{\bf A}(x,y) = {\bf A}_0(y) +
\sum_{m=0}^\infty{(-x)^{m+1}\over (m+1)!}
\ad{R$_{00}$}^m{\bf A}_1(y)\,,&\hbox{~with} \ {\bf A}_1(y)
\equiv {\bf R}_{10} + [{\bf R}_{00},\,{\bf A}_0(y)]\,,\cr
{\bf R}(u,v)  = {\bf R}_0(v) + \sum_{k=0}^\infty{(+u)^{k+1}\over
(k+1)!}\ad{A$_{00}$}^k{\bf R}_1(v)\,,&\hbox{~with}\ {\bf R}_1(v)
\equiv {\bf A}_{10}
+ [{\bf A}_{00},\,{\bf R}_0(v)]\,,\cr}\eqno(A2.2)$$
along with a collection of requirements on the commutators of these vector
 fields, which are most easily
expressed by setting ${\bf A}_{m+1}(y)$
as the $(m+1)$-st term in the expansion, in powers of $x$,
of ${\bf A}(x,y)$, above, and ${\bf R}_{k+1}(v)$ as the $(k+1)$-st term in
the expansion, in powers of $u$, of ${\bf R}(u,v)$
The entire collection of constraints is then easily stated as the
quadruply countable set:
$$ \bigl[{\bf A}_{m+1}(y)\,,{\bf R}_{k+1}(v)\bigr] = 0 \quad\forall\,k,m =
0,1,2,\,\ldots\quad.\eqno(A2.3)$$
\v
We now apply this lemma to the two appropriate equations, in
\Es{9}.  The second of these equations involves ${\bf B}(u,y)$ and
${\bf Z}(x,y)$, so that the $y$-dependence overlaps, but is ``irrelevant''
for this particular pde.  Choosing to use only ${\bf X}_{10}$,
our lemma provides us with
new vector fields, ${\bf R}(y)$, ${\bf X}_0(y)$, and ${\bf X}_1(y)$, such that
$$\eqalign{{\bf B}(u,y) = &\,e^{u\ads{X$_0$}}\,{\bf R}(y) +
\sum_{m=0}^\infty{u^{m+1}\over (m+1)!}\ad{X$_0$}^m\,{\bf X}_1(y)
\quad,\cr{\bf Z}(x,y) = &\,{\bf X}_0(y) + \sum_{\ell=0}^\infty{(-x)^{\ell+1}
\over (\ell+1)!}(\ad R)^\ell\,{\bf X}_1(y)\quad,\cr}\eqno(A2.4)$$
\no along with the commutator requirements that
$$ \bigl[\ad{ R}^{\ell+1}{\bf X}_0\,,\,\ad{X$_0$}^m
({\bf X}_1+[{\bf X}_0\,,\,{\bf R}])\bigr]\quad,\quad\forall\,\ell,m =
0,1,2,\ldots\quad.\eqno(A2.5)$$
\no The same lemma applied to the first equation, involving
${\bf C}(u,x)$ and ${\bf Z}(y,x)$, gives us the existence of
vector fields ${\bf S}(x)$, ${\bf Y}_0(x)$ and ${\bf Y}_1(x)$ such that
$$\eqalign{{\bf C}(u,x) = &\,e^{-u\ads{Y$_0$}}\,{\bf S}(x)
+ \sum_{n=0}^\infty
{(-u)^{n+1}\over (n+1)!}\ad{Y$_0$}^m\,{\bf Y}_1(x)\quad,\cr
{\bf Z}(y,x) = &\,{\bf Y}_0(x) + \sum_{\ell=0}^\infty{(-y)^{\ell+1}\over
(\ell+1)!}\ad{S}^\ell\,{\bf Y}_1(x)\quad,\cr}\eqno(A2.6)$$
\no along with the commutator requirements that
$$ \bigl[\ad{S})^{\ell+1}\,{\bf Y}_0\,,\,\ad{Y$_0$}^n\,
({\bf Y}_1+[{\bf Y}_0\,,\,{\bf C}_0])\bigr]\quad,\quad\forall\,\ell,n =
0,1,2,\ldots\quad.\eqno(A2.7)$$
\vskip-8truept\par
 The requirement that ${\bf Z}(x,y)$, as presented in \Es{A2.4} and
(A2.6), should be the same is a very strong constraint on the
underlying vector fields.  A term-by-term comparison is, in
principle, required.  For instance, the lowest-order requirement is
that ${\bf Y}(x)$ and ${\bf X}(y)$ should be related as follows:
$${\bf Y}_0(x)\, = {\bf Z}_0 + \sum_{\ell=0}^\infty{(-x)^{\ell+1}
\over (\ell+1)!}\ad {R$_{00}$}^\ell\,{\bf X}_{10}\,,\quad
{\bf X}_0(y)\, = {\bf Z}_0 + \sum_{\ell=0}^\infty{(-y)^{\ell+1}\over
(\ell+1)!}\ad {S$_{00}$}^\ell\,{\bf Y}_{10}\,,\eqno(A2.8)$$
\no while we could also write out the requirements on ${\bf Y}_1(x)$,
${\bf X}_1(y)$, etc.  These requirements would then have to be inserted
into the shapes for ${\bf B}$ and ${\bf C}$.
While we have indeed written out yet much more general series of
equations for this problem, we nonetheless feel justified at this point
to append to these equations some additional assumptions that simplify
the problem enough to be presented with only a finite amount of
formalism.
Therefore, at this point, we consider only the case when
${\bf Z}$ is completely independent of both $x$ and $y$, thus
reducing all the expressions for ${\bf Z}$ above to a single term, which
we will refer to as ${\bf Z}_0$, a vector field defined only over the
fiber variables, $w^A$.  As well, we again {\it assume} that it is reasonable
to truncate the series for ${\bf R}(y)$ and ${\bf S}(x)$ to
make them first-order polynomials:
$${\bf R}(y) = {\bf R}_0 + y{\bf R}_1 \quad,\quad
{\bf S}(x) = {\bf S}_0 + x{\bf S}_1 \quad,\eqno(A2.9)$$
\vskip-12truept\par
Insertion of \Es{A2.4, A2.6} and (A2.9) into the remaining portion of
\Es{9} gives us a collection of requirements on the vector fields already
named, i.e., $\{{\bf Z}_0, {\bf R}_0, {\bf R}_1, {\bf S}_0,
{\bf S}_1\}$, as well as additional ones which involve repeated commutators
 with ${\bf Z}_0$ that have not yet been named.  We first list all
the requirements that follow when one evaluates at $u=0$,
therefore involving only those vector fields just named above:
 $${\vbox{\offinterlineskip \halign{ &\vrule#& \dstrut\hfill\ #$\>$\hfill\cr
\omit&\omit&\multispan{11}\hrulefill\cr
 \omit &\omit &width2pt& ${\bf R}_0$ && ${\bf R}_1$
 && ${\bf S}_0$ &&${\bf S}_1$ &&${\bf Z}_0$&\cr
\noalign{\hrule height2pt}\cr
& ${\bf R}_0$ &width2pt&0&&\omit&& ${\bf R}_1-{\bf S}_1$&& ${\bf Z}_0$
&&$+{\bf T}_0$&\cr\tablerule
& ${\bf R}_1$ &width2pt&\omit&&0&& ${\bf Z}_0$ &&0&&$+{\bf V}_0$&\cr
\tablerule
& ${\bf S}_0$ &width2pt& \omit&& \omit&&0&&\omit&&$-{\bf U}_0$&\cr
\tablerule
&${\bf S}_1$ &width2pt&\omit&&\omit&&\omit&&0&&$-{\bf W}_0$&\cr
\tablerule
&${\bf Z}_0$ &width2pt&\omit&&\omit&&\omit&&\omit&&0&\cr
\noalign{\hrule} \cr }}}\eqno(A2.10)$$
\no The quantities in the last row and column are new quantities, that
will be needed at higher powers in $u$, which we now define generically,
 $\forall n = 0,1,2,\,\ldots$~:
$$\matrix{
{\bf T}_n\equiv(-1)^n\ad{Z$_0$}^n[{\bf Z}_0\,,\,{\bf R}_0]\,,\,&
{\bf V}_n\equiv(-1)^n\ad{Z$_0$}^n[{\bf Z}_0\,,\,{\bf R}_1]\cr
\noalign{\vskip6truept}
{\bf U}_n\equiv\ad{Z$_0$}^n[{\bf Z}_0\,,\,{\bf S}_0]\,,\,&
{\bf W}_n\equiv\ad{Z$_0$}^n[{\bf Z}_0\,,\,{\bf S}_1]\cr
\noalign{\vskip6truept}
{\bf X}_{\ell p}\equiv \bigl[{\bf T}_\ell,\,{\bf V}_p\bigr]\,,\,&
{\bf Y}_{mn}\equiv\bigl[{\bf U}_m,\,{\bf W}_n\bigr]\cr
}\quad.\eqno(A2.11)$$
\no The quantities in the last line are not determined by the requirements
of the equation; however, we will show below that it does not permit
them to vanish, so that we now give them names.
\v
The rest of \Es{9} requires two additional sets of commutators
involving these new quantities. The first set involves the
mixed commutators
$$\bordermatrix{&{\bf T}_n&{\bf V}_n&{\bf X}_{np}&{\bf U}_n&{\bf W}_n&{\bf
Y}_{np}\cr
{\bf Z}_0&-{\bf T}_{n+1}&-{\bf V}_{n+1}&{\bf X}_{n+1,p}+
{\bf X}_{n,p+1}&+{\bf U}_{n+1}&+{\bf W}_{n+1}&-{\bf Y}_{n+1,p}-{
\bf Y}_{n,p+1}\cr
{\bf R}_0&&&&-{\bf W}_n&{\bf Z}_0\cr
{\bf R}_1&&&&{\bf Z}_0&0&{\bf W}_{p+1}\cr
{\bf S}_0&-{\bf V}_n&{\bf Z}_0\cr
{\bf S}_1&{\bf Z}_0&0&-{\bf V}_{p+1}\cr}\>,\eqno(A2.12)$$
\no while the
second set describes the commutators between the higher-level ones,
themselves:
$$\bordermatrix{&{\bf T}_n&{\bf V}_n&{\bf U}_n&{\bf W}_n&{\bf X}_{np}&
{\bf Y}_{np}\cr
{\bf T}_m&&{\bf X}_{mn}&0&-{\bf Z}_0&&+{\bf U}_{n+1}\cr
{\bf V}_m&-X_{nm}&&-{\bf Z}_0&0&&-{\bf W}_{p+1}\cr
{\bf U}_m&0&+{\bf Z}_0&&{\bf Y}_{mn}&+{\bf T}_{n+1}\cr
{\bf W}_m&+{\bf Z}_0&0&-{\bf Y}_{nm}&&-{\bf V}_{p+1}\cr
{\bf X}_{mk}&&&-{\bf T}_{m+1}&+{\bf V}_{k+1}&&2{\bf Z}_0\cr
{\bf Y}_{mk}&-{\bf U}_{m+1}&+{\bf W}_{k+1}&&&-2{\bf Z}_0\cr}\>,
\eqno(A2.13)$$
\v
The structure above is of course still frightfully complicated;
therefore one surely wonders how much of it is ``necessary.''
The Lagrange multipliers supply the answer to at least part of that
question.  Referring
back to \Es{7}, within the current notation they are simply the
three vector fields
$\{{\bf Z}_0\,,\, {\bf T}_0 + y\,{\bf V}_0\,,\, {\bf U}_0 + x\,{\bf W}_0\}$.
Considering that $\bigl[{\bf X}_{mk},\,{\bf Y}_{np}\bigr] = +2\,{\bf Z}_0$,
independent of the
values of the indices $\{m,k,n,p\}$, we immediately see that
none of those undetermined double commutators
${\bf X}_{mk}$, nor ${\bf Y}_{np}$, may vanish.
  However, if any of the individual terms within our three vector
fields were to vanish, then
one or more of these double commutators would indeed have to vanish,
therefore requiring us to maintain, at the least, all 5 of those
vector fields non-zero and linearly independent.
\v
As a first approach to studying this structure, we
go to a very simplified homomorphic image which we name ${\cal{RT}}_0$.
The mapping is generated by
dropping the subscripts on the newly-created quantities:
$$\eqalign{&\matrix{
{\bf T}_n\longrightarrow {\bf T}_0\equiv {\bf T}\,,&
{\bf V}_n\longrightarrow {\bf V}_0\equiv {\bf V}\cr
\noalign{\vskip6truept}
{\bf U}_n\longrightarrow {\bf U}_0\equiv {\bf U}\,,&
{\bf W}_n\longrightarrow {\bf W}_0\equiv {\bf W}\cr}\quad, \quad\forall n =
0,1,2,\,\ldots\,,\cr
&\hskip1,2truein {\bf Z}_0 \longrightarrow {\bf Z}\,,\cr}
\eqno(A2.14)$$
\no where ${\bf Z}_0\rightarrow {\bf Z}$ is just to make
the typography all appear more consistent.  (This is also a useful place to
point out that if the second-order terms in $x$ or $y$ had been kept,
all their commutators would now be either zero or undetermined, thus
motivating our having already dropped them.)
We should perhaps also mention that a somewhat
more complicated mapping does not work, namely one where ${\bf T}_n$
might have been mapped to $(a_{\scriptscriptstyle T})^n
{\bf T}$, for some constant $a_{\scriptscriptstyle T}$.
In fact, such quantities $a_i$ ~are completely determined by the
requirement that this
mapping actually be a homomorphism.)
\v
Our algebra ${\cal RT}_0$ is generated by
$\{{\bf Z},{\bf R}_0,{\bf R}_1,{\bf T},{\bf V},{\bf S}_0,{\bf S}_1,
{\bf U},{\bf W},
{\bf X},{\bf Y}\}$, and the new Lie product table is just obtained by
ignoring the subscripts in the previous tables.
Using those tables, we note the existence of a very interesting
subalgebra contained within ${\cal{RT}}{}_0$, namely the one where
we drop out the $\{{\bf R}_i\,,{\bf S}_j\}$.  This subalgebra is
generated by $\{{\bf Z}, {\bf T}, {\bf V}, {\bf U}, {\bf W}\}$, and we will
refer to it as ${\cal {RT}}{}_{00}$.
Remembering that it will also have $\bf X$ and $\bf Y$ as elements,
the appropriate commutator table is\par
$${\cal {RT}}{}_{00}\,:\qquad\bordermatrix{
&{\bf Z}&{\bf U}&{\bf W}&{\bf T}&{\bf V}&{\bf X}&{\bf Y}\cr
{\bf Z}&0&{\bf U}&{\bf W}&-{\bf T}&-{\bf V}&-2\,{\bf X}&+2\,{\bf Y}\cr
{\bf U}&&0&{\bf Y}&0&{\bf Z}&{\bf T}\cr
{\bf W}&&&0&{\bf Z}&0&-{\bf V}\cr
{\bf T}&&&&0&{\bf X}&&{\bf U}\cr
{\bf V}&&&&&0&&-{\bf W}\cr
{\bf X}&&&&&&0&+2\,{\bf Z}\cr
{\bf Y}&&&&&&&0\cr}\>.\eqno(A2.15)$$
\no Since none of these elements is allowed to
vanish, this is the fundamental
subalgebra within our general prolongation structure.
It is isomorphic to the contragredient algebra
of infinite growth, $K_2$, described in the main text, as given by
$${\bf h} \longrightarrow {\bf Z}\>,\;
{\bf e}_1 \longrightarrow {\bf U} \quad,\quad {\bf e}_2
\longrightarrow {\bf W}\quad ,
\quad{\bf f}_1 \longrightarrow {\bf V} \quad,\quad
{\bf f}_2 \longrightarrow {\bf T}\quad .\eqno(A2.16)$$
\v
The subalgebra ${\cal {RT}}{}_{00}$,
isomorphic to $K_2$, is however {\bf not} the
entirety of ${\cal {RT}}{}_{0}$, which also0 contains
the generators ${\bf R}_i$ and ${\bf S}_j$.  It is straight-forward to
show that simply mapping them to zero will not work; i.e., the
requirements of the Jacobi identity will cause the structure
to collapse sufficiently far that it ``forgets'' the original pde.
In an attempt to better understand the function of these objects, we look
for linear combinations of them which are ``eigenvectors'' of ${\bf Z}$,
and find that the four new combinations
${\bf A}\equiv {\bf R}_0 + {\bf T}$, ${\bf B}\equiv {\bf S}_0-{\bf U}$,
${\bf M}\equiv \bf R_1 + {\bf V}$, and ${\bf N}\equiv {\bf S}_1 - {\bf W}$ all
commute
with ${\bf Z}$. The requirements of the Jacobi
identity and {\bf also} the requirements of the Lagrange multipliers
do not prevent us from setting $\bf M$ and $\bf N$ to 0, which we
therefore do.  On the other hand, taking the remaining two, $\bf A$,
and $\bf B$,
as generators in lieu of the original $\bf R_0$ and $\bf S_0$,
respectively, we find that $\bf A + B$ is a
central element, while neither $\bf A$ nor $\bf B$
occur in the commutator
ideal.  We may therefore rescue the ``rest'' of the algebra by ignoring
that central element, and viewing this algebra as the {\it
semi-direct
product} of $K_2$ and the algebra consisting of the single element $\bf A$.
We maintain the mapping
as given in \E{A2.16} and append to it the commutation relations with
$\bf A$, as follows:
$$ [{\bf A}, {\bf h}] = 0\,,\;[ {\bf A},{\bf e}_1] = 0\,,\;[{\bf A},
{\bf e}_2] = -{\bf e}_1
\,,\;[{\bf A},{\bf f}_1] = +{\bf f}_2\,,\;[{\bf A},{\bf f}_2] =
0\quad,\eqno(A2.17)$$
\no and can present the prolongations of the total derivatives in the
form
$${\bf F} = p{\bf h} + {\bf A} - e^{-u}({\bf f}_1 + y\,{\bf f}_2)
\;,\quad{\bf G} =
-q{\bf h} - {\bf A} + e^{-u}({\bf e}_2 + x{\bf x}_1)\quad,\eqno(A2.18)$$
\no which should be compared, for instance, with the result in the main
text, at \Es{17}.
\v The form of $\bf F$ and $\bf G$ given in \Es{A2.18} is clearly
different from that in \Es{17}; One would argue that this is not
surprising since they correspond to two distinct sets of assumptions
concerning the dependence of the prolongation quantities on the independent
variables.  However, it turns out that the two sets are in fact
equivalent under a gauge transformation.  From the viewpoint of \E{3},
the quantities $\bf F$ and $\bf G$ are prolongations of the total
derivatives on $J^\infty$ to the (larger) covering
space $J^\infty\otimes W$.  Moreover, the so-prolonged derivatives only
commute when one restricts the calculation to the subvariety defined by
the pde being studied.  Therefore it is reasonable to treat the quantities
$F^A\,dx + G^A\,dt$ as (the coefficients of a Lie-algebra-valued) connection
1-form on the covering space.  Therefore, it should transform in the usual
way for connections.  The transformations we want to consider correspond
to flows of the covering space generated by particular tangent vector fields,
so that we are simply moving along a congruence of curves.  Moreover, since
we are restricting our attention to vertical vector fields, different
values of a parameter along the curves just correspond to different
choices for values of the fiber coordinates over the same base point.
(See Refs. 22 or 25 for considerably more discussion concerning this idea.)
The structure of our theory should be independent of distinctions such
as this; therefore, we refer to transformations of this type, which
simply map different explicit presentations of the underlying geometry
into one another, as $\bf gauge$ transformations.  Given a vertical
vector field, $\bf R$ defined over some (local) portion of our manifold,
the flow of that vector field is a (local) mapping of the manifold into
itself, that can be presented via a congruence of curves, described by
$\Phi_t \equiv e^{t{\bf R}}:\,U\subseteq M \rightarrow M$.  Under the
induced mapping of the tangent bundle, the transformation law for
an arbitrary connection 1-form, $\Gamma$, would be
$$ \Gamma' \equiv \Gamma_t = e^{t\ads{R}}\Gamma - d(t{\bf R})
\quad.\eqno(A2.19)$$
\v
For our particular transformation, we will choose the vector field $\bf R$,
above to be our ``extra''
algebra element, $\bf A$, and the flow parameter, $t$, as $x-y$.
Performing the transformation
on the $\bf F$ and $\bf G$ given in \Es{A2.18}, we find that the so-transformed
quantities, $\bf F'$ and $\bf G'$ are in fact identical to the ones given
in \Es{17}:
$$\eqalign{{\bf F}' & \equiv e^{(x-y)\ads{A}}{\bf F} - D_x\{
(x-y){\bf A}\} \quad = \quad +p\,{\bf h} - e^{-u}({\bf f}_1 + x{\bf
f}_2)\quad,\cr
{\bf G}' & \equiv e^{(x-y)\ads{A}}{\bf G} - D_y\{
(x-y){\bf A}\} \quad = \quad -q\,{\bf h} + e^{-u}({\bf e}_2
+ y{\bf e}_1)\quad.\cr}\eqno(A2.20)$$
\no Since
$\bf A$ is a vertical vector field, this transformation simplify
re-defines the
origin in our fiber spaces in a manner that depends explicitly upon the
value of the independent variable $x-y$.  This transformation has two
immediate effects.  It was chosen to remove the vector
field $\bf A$ from the presentation for $\bf F$ and $\bf G$.  As well,
it has switched the $x$- and $y$-dependence of $\bf F$ and $\bf G$.
Of course an arbitrarily chosen dependence of the connection on $\bf A$
would {\bf not} have allowed one to remove it.  That this was possible
shows that the explicit
dependence of $\bf F$ and $\bf G$ on the independent variables was in
fact gauge-dependent.

\vfill\eject
\centerline{REFERENCES}
\parindent=0pt
\baselineskip=14pt
\def\hi{\hangindent=20truept}
\def\bn#1{{\bf #1}}

\frenchspacing
\v
\hi ~1.  Ivor Robinson and Andrezj Trautman, ``Some spherical gravitational
waves in general relativity,'' Proc. R. Soc. A{\bf 265}, 463-473 (1962).

\hi ~2.  Piotr T. Chrusciel, ``On the global structure of Robinson-Trautman
space-times,''
Proc. R. Soc. {\bf A436}, 299-316 (1992), ``Semi-global Existence and
Convergence of Solutions of the Robinson-Trautman (2-Dimensional Calabi)
Equation,''
Commun.Math. Phys. \bn{137}, 289-313 (1991); J. Foster and E. T. Newman,
``Note on the Robinson-Trautman Solutions,''
J. Math. Phys. \bn{8}, 189-94 (1967).

\hi ~3.  Dietrich Kramer, Hans Stephani, Eduard Herlt, and Malcolm MacCallum,
{\it Exact Solutions of Einstein's field equations}, Cambridge Univ. Press,
Cambridge, UK, 1980.

\hi ~4.  Hans Stephani, {\it Differential equations:  Their solution using
symmetries}, Cambridge University Press, Cambridge, UK, 1989.  All the Lie
symmetries are determined here, and discussed.  Various people have made
searches (unpublished) for generalized symmetries, all unsuccessful.

\hi ~5.  Frank Estabrook and Hugo Wahlquist, ``Prolongation structures of
nonlinear evolution equations,'' J. Math. Phys. {\bf 16}, 1-7
(1975), and Hugo Wahlquist and Frank Estabrook, ``Prolongation structures
of nonlinear evolution equations. II.'' J. Math. Phys. {\bf 17},
1293-7 (1976).

\hi ~6.  I. S. Krasil'shchik and A. M. Vinogradov, ``Nonlocal Symmetries
and the Theory of Coverings:  An Addendum to A. M. Vinogradov's Local
Symmetries and Conservation Laws,'' Acta
Applicandae Mathematicae {\bf 2}, 79-96 (1984); I. S. Krasil'shchik and
A.M. Vinogradov, ``Nonlocal Trends in the Geometry of Differential
Equations:  Symmetries, Conservation Laws, and B\"acklund Transformations,''
Acta Applicandae Mathematicae {\bf 15}, 161-209 (1989).

\hi ~7.  Hugo Wahlquist and Frank Estabrook,  ``B\"acklund Transformation
for Solutions of the Korteweg-de Vries Equation,''
Phys. Rev. Lett. \b{31}, 1386-90 (1973).

\hi ~8.  C. Hoenselaers, ``More Prolongation Structures,'' Progress of Theor.
Physics \b{75}, 1014-29 (1986),
and ``Equations admitting O(2,1)$\times$R(t,t$^{-1}$) as a prolongation
algebra,'' J. Phys. A \b{21}, 17-31 (1988).

\hi ~9.  J.H.B. Nijhof and G.H.M. Roelofs, ``Prolongation structures of a
higher-order nonlinear Schr\"odinger equation,'' J. Phys. \bn{A25},
2403-16 (1992), G. H. M. Roelofs and R. Martini, ``Prolongation structure of
the KdV equation in the bilinear form of Hirota,''
J. Phys. \bn{A23}, 1877-84 (1990),
W. M. Sluis and P.H.M. Kersten, ``Non-local higher-order symmetries for the
Federbush model,'' J. Phys. \bn{A23}, 2195-2204 (1990).

\hi 10.  B. Kent Harrison, ``Unification of Ernst-equation B\"acklund
transformations using a modified Wahlquist-Estabrook technique," J. Math.
Phys. \b{24} 2178-87 (1983).

\hi 11.  E.N. Glass \& D.C. Robinson, ``A nilpotent prolongation of the
Robinson-Trautman equation,''
J. Math. Phys. \b{25}, 3382-6 (1984).

\hi 12.  A. W.-C. Lun and Cornelius Hoenselaers, reported at the 1993 Jena
Relativit\"atstheorie Conference, unpublished.

\hi 13.   I. M.
Gel'fand and A. A. Kirillov, {\it Sur les corps li\'es aux alg\'ebres
enveloppantes des alg\'ebres de Lie}, Inst. Hautes
\'Etudes Sci. Publ. Math. {\bf 31}, 5-19 (1966).

\hi 14.  V. G. Kac, ``Simple Irreducible Graded Lie algebras of finite
growth," Math. USSR-Izvestija {\bf 2}, 1271-1311 (1968).

\hi 15.  V. G. Kac, ``On Simplicity of certain Infinite Dimensional Lie
Algebras," Bull. Amer. Math. Soc. {\bf 2}, 311-4 (1980).

\hi 16.   V.E. Zakharov and A.B. Shabat, ``A Scheme for Integrating the
Nonlinear Equations of Mathematical Physics by the Method of the Inverse
Scattering Problem I.''
Functional Analysis \& Applications \bn8, 43-53 (1974), and
``Integration of Nonlinear Equations of Mathematical Physics by the
Method of Inverse Scattering. II'' \bn{13}, 13-22 (1976), and
``Exact Theory of Two-Dimensional Self-Focusing and One-Dimensional
Self-Modulation of Waves in Nonlinear Media,'' J.E.T.P. \bn{34}, 62-9 (1972).

\hi 17.  Elie Cartan, {\it Les syst\`emes diff\'erentiels ext\'erieurs et
leurs applications g\'eom\'etriques}, Hermann, Paris, 1945.

\hi 18.  J. D. Finley, III and John K. McIver, ``Prolongations to Higher
Jets of Estabrook-Wahlquist Coverings for PDE's''
Acta Applicandae Mathematicae \bn{32}, 197-225 (1993).

\hi 19.  P. Molino, ``General prolongations and (x,t)-depending
pseudopotentials for the KdV equation,''
J. Math. Phys. \b{25}, 2222-5 (1984), discusses why this
can be done for the KdV equation in a way which easily generalizes to the
case when the independent variables do not appear explicitly in the equation,
and that equation is quasilinear.

\hi 20.  A. N. Leznov, M. V. Saveliev, {\it Group-Theoretical Methods for
Integration of Nonlinear Dynamical Systems}, Birkh\"auser Verlag, Basel, 1992.

\hi 21.  F. Pirani, D. Robinson and W. Shadwick, {\it Local Jet Bundle
Formulation of B\"acklund Transformations\/}, Mathematical Physics
Studies, Vol. 1, Reidel, Dordrecht, 1979.

\hi 22.  J. D. Finley, III and John K.
McIver, ``Infinite-Dimensional Estabrook-Wahlquist
Prolongations for the sine-Gordon Equation,"
J. Math. Phys., to be published.

\hi 23.  W.F. Shadwick, ``The B\"acklund problem for the equation
$d^2\,z/dx^1\,dx^2 = f(z)$,'' J. Math. Phys. \bn{19}, 2312-2317 (1978).

\hi 24.  C. Rogers and W. F. Shadwick, {\it B\"acklund Transformations and
Their Applications\/}, Academic Press, New York, 1982.

\hi 25.  L. O'Raifeartaigh, {\it Group structure of gauge theories},
Cambridge Univ. Press, Cambridge, UK, 1986.

\vfill\eject
\bye